\documentclass[pra, superscriptaddress, showpacs, amsmath,amssymb, aps, 11pt]{revtex4-1}

\usepackage{amsmath,amsthm}
\usepackage{amssymb}
\usepackage{amscd,bm}
\usepackage{wasysym}
\usepackage[ansinew]{inputenc}
\usepackage[T1]{fontenc}
\usepackage{ae,aecompl}
\usepackage{hyperref}

\usepackage[rightcaption]{sidecap}

\usepackage[pdftex]{graphicx}
\usepackage{braket}

\newcommand{\ud}{\mathrm{d}}
\newcommand{\ic}{\mathrm{i}}

\newcommand{\zZ}{\mathbb{Z}}
\newcommand{\sgn}{\mathrm{sgn}}

\begin{document}

\title{Applications of fidelity measures to complex quantum systems}

\author{Sandro Wimberger}
\affiliation{Dipartimento di Fisica e Scienze della Terra, Universit\`a di Parma, Via G.P. Usberti 7/a, I-43124 Parma}
\affiliation{INFN, Sezione di Milano Bicocca, Gruppo Collegato di Parma, Italy}
\affiliation{Institut f\"ur Theoretische Physik, Philosophenweg 12, Universit\"at Heidelberg, D--69120 Heidelberg}



\email{sandromarcel.wimberger@unipr.it}

\date{\today }

\begin{abstract}
We revisit the fidelity as a measure for the stability and the complexity of the quantum motion of single and many-body systems. Within the context of cold atoms, we present on overview of applications of two fidelities which we call static and dynamical fidelity, respectively.  The static fidelity applies to quantum problems which can be diagonalized since it is defined via the eigenfunctions. In particular, we show that the static fidelity is a highly effective practical detector of avoided crossings characterizing the complexity of the systems and their evolutions. The dynamical fidelity is defined via the time-dependent wave functions. Focussing on the quantum kicked rotor system, we highlight a few practical applications of fidelity measurements in order to better understand the large variety of dynamical regimes of this paradigm of a low-dimensional system with mixed regular-chaotic phase space.
\end{abstract}

\maketitle

\section{Introduction}
\label{sec:1}

Generally it is very useful to have a working tool for the characterization of systems and their dynamical evolution. An experimental technique with great value for practical measurement applications is the echo method, in particular known as spin or Hahn echo and much used in nuclear magnetic resonance experiments \cite{hahn}. Echos are necessarily dynamical functions characterizing the quality and the difference between forward and backward evolutions. As a consequence, it was quite natural to suggest the echo in the context of dynamical systems as a measure for the dynamical spreading of trajectories, which was done by Asher Peres back in 1984 \cite{peres}. This quantity is defined as the overlap of two quantum mechanical wave function developed with two slightly different Hamiltonians, whose difference is quantified by the change of an appropriate control parameter. While it turned out that a categorization of fidelity decays for different dynamical systems, with classically integrable, mixed regular-chaotic or fully chaotic phase space, is generally not simple and may depend on non universal details of the systems, it is well established that several universal decay regimes can be found characterized not so much by the systems' properties but by the perturbation strength. These universal regimes are known as perturbation regime, Gaussian decay and Fermi Golden Rule regime \cite{JP2001} and they are well described in recent reviews \cite{review,Prosen}.

In this paper we present a compact overview of the use of the overlap function, vulgo fidelity, in the broader context of complex quantum systems, which range from quantum many-body models \cite{PLW2011} to strongly driven dynamical quantum systems \cite{SW2011}. We review two concepts of overlap functions relevant for cold-atom experiments and which both are equally relevant for characterizing the response of quantum systems to perturbations as well as their dynamical evolution. The first concept is based on the static overlap of two eigenfunction computed for two slightly different control parameters, hence it is basically a spectral characterization of the system at hand. The second concept is the original fidelity introduced by Peres \cite{peres}, which represents the overlap of two dynamical wave functions taken at equal times but evolved with two different Hamiltonians up to this time. Both measures are useful to characterize the temporal evolution of a problem, since the latter is connected with the spectrum by the spectral theorem, of course. What we mean is that knowing all static overlaps between eigenfunctions at different control parameters and their corresponding eigenvalues, we can in principle reconstruct the dynamical overlap function. This is true for Hamiltonians which either do not depend explicitly on time or are periodically time-dependent (since then we can use Floquet theory to arrive at an eigenbasis). The advantage of the purely dynamical overlap function is that it can be computed also for Hamiltonians which  are explicitly time dependent, and there is no need for computing eigenvalues and eigenfunctions, which can be a true practical problem for large quantum few-body or many-body systems \cite{javier,PMW2015}.

\section{Static fidelity measure}
\label{sec:2}

Given a control parameter $\lambda$ and a perturbation $V$, we are studying the following class of Hamiltonians $H(\lambda) = H_0 + \lambda V$. Then the static fidelity between the $n$-th eigenstates (assuming normalized states, a discrete spectrum, and an ordered spectrum such as $E_{n+1} \geq E_n$), denoted by $|n\rangle$, of the two Hamiltonians $H(\lambda)$ and $H(\lambda + \delta\lambda)$ is defined as 
\begin{equation}\label{eq:1}
f_n(\lambda, \delta\lambda) \equiv \lvert\langle n(\lambda) | n(\lambda+\delta\lambda)\rangle\rvert \,.
\end{equation}
Here the change in the control parameter $\delta\lambda$ is assumed to be small. Locally in the spectrum, the quantity $f_n(\lambda,\delta\lambda)$ characterizes the change of the eigenfunctions. If the latter change adiabatically, the value of $f_n(\lambda,\delta\lambda)$ will stay close to one. On the other hand, if we encounter avoided level crossings, at which the eigenfunctions strongly mix, the value of $f_n(\lambda,\delta\lambda)$ will substantially decrease.  To better detect and characterize avoided crossings for a given quantum level $n$, 
we better investigate the fidelity change~\cite{PLW2011}, also known as fidelity susceptibility in condensed-matter literature \cite{Ref1}, defined by
\begin{equation}\label{eq:fidchange}
		S_n(\lambda, \delta\lambda) \equiv \frac{1 - f_n(\lambda, \delta\lambda) }{(\delta\lambda)^2}  \,.
\end{equation}
For most of the spectra so far encountered in practice, as long as $\delta\lambda\ll1$, $S_n$ turned out to be largely independent of $\delta\lambda$, i.e. $S_n(\lambda, \delta\lambda)  \approx S_n(\lambda)$,  and vanishingly small except in the vicinity of an avoided level crossing. The independence of
$\delta\lambda$ follows in the limit $\delta\lambda\to 0$ directly from perturbation theory.

The independence of $\delta\lambda$ has a simple origin, namely that the first non-vanishing contribution to $f_n$ in the perturbation expansion of the changed state $|n(\lambda +\delta\lambda )\rangle$ is of second order in $\delta\lambda$, a well known fact from quantum perturbation theory. The fidelity measure (\ref{eq:1}) and the fidelity susceptibility (\ref{eq:fidchange}) had found a nice application in the characterization of {\em ground states} of quantum many-body systems \cite{BV2007,Ref2,Ref1}. $S_n$ detects qualitative changes in the ground state very well, when they are connected to an avoided level crossing. Since most of quantum phase transitions in finite systems are characterized by a gap, $S_n$ acts as an effective detector of these transitions and one can check its scaling with the system size, for instance, to pin down the interesting parameter value at which the transitions occurs. Our definition above generalizes the utility of $S_n$ to gaps, e.g. avoided crossings, lying anywhere in the spectrum. 

\begin{figure}[t!]
	\begin{center}
	\includegraphics[width = 0.75\linewidth]{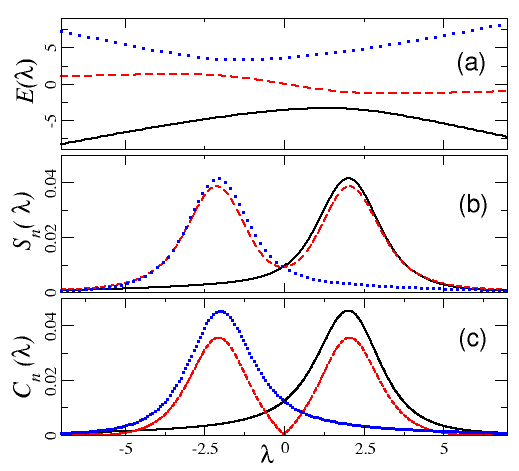}
	\caption{\label{fig:1} (a) Energy spectrum of Eq.~(\ref{eq:tripleAC}) for $x=0, y =2, z=3$. All levels are coupled and the spectrum shows two close avoided crossings; (b) fidelity change $S_n(\lambda)$ and (c) renormalized curvature $C_n(\lambda)$ for the three levels shown in (a). Data adapted from \cite{PLW2011}.}
	\end{center}
\end{figure}

Of course, the computation of $S_n$ is based on the eigenfunctions, which is a practical disadvantage for large quantum systems, where its simpler to compute just the eigenvalues (and not the eigenfunctions explicitly) with optimized numerical methods. Yet this problem turns into an advantage in the sense that it turned out that $S_n$ detects a large percentage, close to one hundred per cent, of the total number of avoided crossings in either a region of the spectrum or the entire spectrum of complex quantum systems. Hence, compared to other measures, e.g. which are based solely on the eigenvalues and their change with the control parameter, our measure $S_n$ is very effective and reliable for the practical numerical detection of avoided crossings all over the spectrum. 

\begin{figure}[t!]
	\begin{center}
	\includegraphics[width = 0.9\linewidth]{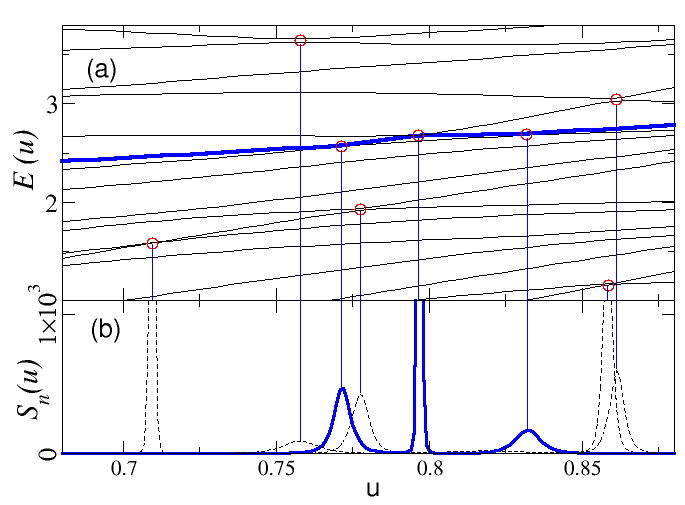}
	\caption{\label{fig:2} (a) Section of the energy spectrum of the Bose-Hubbard model from Eq. (\ref{eq:BHM}) with $F=0$ and for $U = u$ and $J = 1 - u$ for a small system 5 particles in 5 lattice sites. One level undergoing several avoided crossings is highlighted by the thick (blue) line. Some crossings are not resolved on the scale of the figure but marked by circles. (b) Individual fidelity measures $S_n$ of the different eigenstates (dashed lines) and the highlighted level of (a) (thick solid blue line). In this example all avoided crossings are effectively and easily detected.
}
	\end{center}
\end{figure}

There is an intrinsic connection between our measure $S_n$ and the local level curvature in the spectrum. In order to see this connection, let us expand the eigenfunction $|n(\lambda+\delta\lambda)\rangle$ in second order in $\delta\lambda$
\begin{equation}\label{eq:approx}
	S_n(\lambda) \approx \frac{1}{2} \sum_{m\neq n} \frac{\lvert\langle m(\lambda) | V | n(\lambda)\rangle\rvert^2}{[E_n-E_m]^2} \approx \frac{\lvert\langle n'(\lambda) | V | n(\lambda)\rangle\rvert^2}{2\;[E_n-E_{n'}]^2}\,.
\end{equation}
In the last step, we reduced the sum near an isolated avoided crossing to the nearest neighbour level $n'$. Similarly, one can expand the renormalized curvature with the gap function $\Delta(\lambda)=|E_{n}-E_{n'}|$, also in second order to approximate the level curvature by \cite{plotz}
\begin{equation}
\label{eq:curvature}
C_{n}(\lambda) \equiv \bigg|\frac{1}{\Delta(\lambda)}\frac{\partial^2 E_{n}(\lambda)}{\partial\lambda^2}\bigg| \approx
\bigg|\frac{2}{\Delta(\lambda)} \sum_{m\neq n} \frac{\lvert\langle m(\lambda) | V | n(\lambda)\rangle\rvert^2}{E_n-E_m}\bigg| \\ \approx 2\frac{\lvert\langle n'(\lambda) | V | n(\lambda)\rangle\rvert^2}{[E_n-E_{n'}]^2}\,.
\end{equation}
Within this approximation, which is obviously exact for a two level system, we have the following interesting relation between the measure $S_n$ and the level curvature $C_n(\lambda) \approx 4S_n(\lambda)$. This in turn justifies the use of $S_n$ for the numerical detection of avoided crossings, which are intimately connected to changes in the curvature of the eigenvalues as a function of the control parameter. Of course, the last steps in the Eqs. (\ref{eq:approx}) and (\ref{eq:curvature}) are crude approximations. Experience tells that on average over a full spectrum they are quite good though. In any case, our numerical results and computations in general do not make use of them explicitly. From the practical point of view it turns out the detection of avoided crossings by the fidelity is much more reliable than by using the levels. This defines the value of $S_n$ for practical numerical work.

We are testing the measure from Eq. (\ref{eq:fidchange}) in the following. We start with a very simple three state model described by the real symmetric matrix 
\begin{equation}\label{eq:tripleAC}
	H(\lambda) = \left(\begin{array}{ccc} 0 & x & y \\ x & 0 & z \\ y & z & 0 \end{array}\right) + \lambda \left(\begin{array}{ccc} -1 & 0 & 0 \\ 0 & 0 & 0 \\ 0 & 0 & 1 \end{array}\right) = \left(\begin{array}{ccc} -\lambda & x & y \\ x & 0 & z \\ y & z & \lambda\end{array}\right) \,.
\end{equation}
Figure \ref{fig:1} nicely shows that $S_n$ is able to detect and to distinguish two nearby avoided crossings. 

As an example of a quantum many-body system, we study now the class of Hamiltonians of the form
\begin{equation}\label{eq:BHM}	
		\hat{H}(t) = - J \sum_{l=1}^{L} (\text{e}^{i Ft} \hat{a}_{l+1}^{\dagger}\hat{a}_{l} + \text{h.c.}) + U \sum_{l=1}^{L} \hat{n}_l(\hat{n}_l-1) \,. 
\end{equation}
$\hat{a}_{l}^{\dagger}$ and $\hat{a}_{l}$ are creation and annihilation operators of bosons at the lattice site $l$ and $\hat{n}_l=\hat{a}_{l}^{\dagger}\hat{a}_{l}$ is the corresponding number operator. For $F=0$, this model is the well known Hubbard model with onsite interactions (last term) for bosonic ultracold atoms \cite{JZ2005,Kol2004}. Figure \ref{fig:2} shows a scan of the many-body spectrum of a small system of five atoms in five lattice sites as a function of $U\equiv u$, fixing $J=1-u$ and $F=0$. We see that all avoided crossings are detected by our fidelity measure. The next figure \ref{fig:3} shows a histogram count of avoided crossings as a function of $U/J$ now. The zone in parameter space with a large number of crossings corresponds to the spectral region which follows a Wigner-Dyson distribution for the nearest neighbour spacings in the spectrum (one possible definition of quantum chaotic system \cite{haake2010,wimberger2014}). This is shown by the thick/red solid line in the figure. In ref. \cite{lubasch} further details may be found on the connection between quantum chaos, avoided crossings and also entanglement measures in time-independent Bose-Hubbard systems.

\begin{figure}[t!]
	\begin{center}
	\includegraphics[width = 0.9\linewidth]{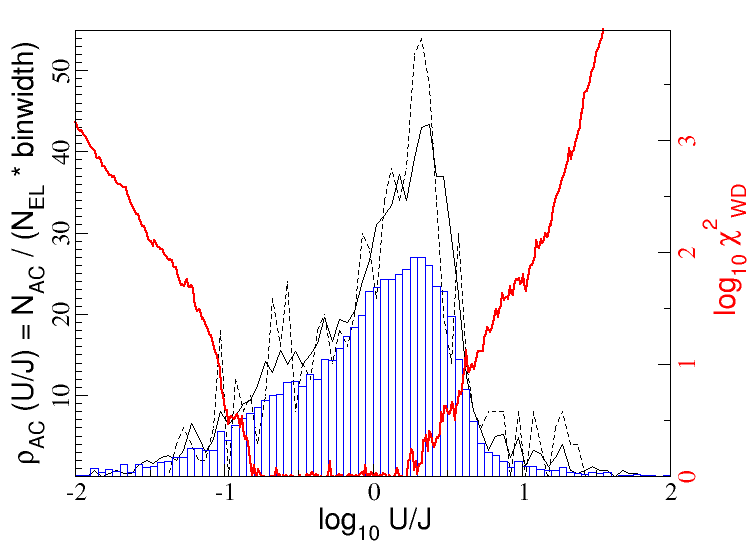}
	\caption{\label{fig:3} (a) Local density of avoided crossings $\rho_{AC}$ (left axis) for $N_{EL}$ energy levels from the centre of the spectrum for the Bose-Hubbard model from Eq. (\ref{eq:BHM}) as a function of U/J for 8 bosons in 7 lattice sites ($F=0$):  $N_{EL}$ = 10 (dashed), 100 (solid) and ca. 400 eigenstates
(blue histogram); a $\chi_2$, see \cite{Tom2008} for details, test is also shown (right axis and thick/red solid line) with values close to zero for good Wigner-Dyson (quantum chaotic) statistics of the full spectrum. The central part of the spectrum follows the overall distribution of crossings, but it has a larger local density of crossings especially around the maximum of the distribution. Interestingly, this maximum is found close to the Mott-insulator transition in the ground state of our 1D Bose-Hubbard model \cite{JZ2005}, a fact that most likely has do with the reordering of the spectrum around this transition briefly discussed in \cite{Kol2004}.
}
	\end{center}
\end{figure}

The case with finite $F$ represents a paradigm of a many-body Floquet Hamiltonian, which is analyzed in detail in references \cite{plotz, PLW2011, PMW2015}. Ref. \cite{PLW2011} shows, in particular, that the transition from the regular to chaotic regime as a function of $F$ can be well captured by the analysis of the width distribution of the detected avoided crossings\cite{ZK1991,PLW2011}
\begin{equation}\label{eq:wd}
	P(c) = (1-\gamma) \delta(c) + \frac{2\gamma^2}{\pi \bar c}\;\text{exp}\left[-\frac{\gamma^2c^2}{\pi \bar c^2}\right]\,.
\end{equation}
We use the fidelity measure $S_n$ to find all avoided crossings and then just compute the difference $c$ of nearest neighbouring levels at the avoided crossings from the spectrum independently. This distribution has a chaotic part of weight $0\leq\gamma\leq1$ and a finite regular component, which is visible as a strong enhancement of $P(c)$ close to zero, see the $\delta (c)$ function. $\bar c$ is the mean value of the width in the ensemble of avoided crossings. $P(c)$ separates better the regular from the chaotic components of a given quantum spectrum than the nearest neighbor level spacing distribution. Using the fidelity change (\ref{eq:fidchange}), the perturbation may or may not preserve symmetries intrinsic to the system. Hence, for an analysis based on our static fidelity measure a splitting of the basis into symmetry reduced subspaces may not be strictly necessary (for symmetry breaking perturbations) in order to characterize the chaotic properties of a quantum system. This is another advantage with respect to standard measure such as the nearest level spacing distributions \cite{haake2010,wimberger2014}.

The static fidelity measure (\ref{eq:fidchange}) is hence a very reliable technique for analysing complex quantum spectra. Its great advantage is that it can easily be automatized and it detects almost all avoided crossings. The measure only fails in cases where two avoided crossings get too close to each other such that their clear separation is not possible any more. Too close means here that their distance in the control parameter becomes comparable to their width in energy.  Such close coincidences of crossings of comparable strength are from our experience rather rare in typical quantum systems. Besides in the mentioned applications for detecting avoided crossings (see above) and, in particular, quantum phase transitions \cite{BV2007}, the authors of \cite{javier} have recently applied our measure to characterize the complexity of an electronic two-body quantum dot systems based on heavy numerical diagonalizations.

We finally note that the concept of the static fidelity measure based on Eq.~(\ref{eq:1}) is easily extended to matrices of overlaps of all eigenfunctions at given control parameters, see e.g. ~\cite{hiller}. The disadvantage is then, however, that one obtains too much information, which may be hard to interpret, and one must compute, of course, all the eigenfunctions of a given system in order to get all the non-diagonal elements of this generalized fidelity matrix.

\section{Dynamical fidelity function}
\label{sec:3}

The stability of quantum evolution against parametric changes is a subject of broad theoretical and experimental interest. 
A widely used concept is the fidelity introduced by Peres \cite{peres}, and the closely related Loschmidt echo \cite{review,Prosen}, 
which is built as the interference pattern between wave functions that are obtained by propagating the {\em same} initial
state under two slightly different Hamiltonians. The behaviour of fidelity, e.g. the quantity
\begin{equation}
  \label{defF}
  F(t)=\left| \langle \psi_{\lambda+\delta \lambda} (t)  | \psi_ {\lambda}(t) \rangle \right|^2 \,,
\end{equation} 
in time is known to display some universal properties that reflect also the underlying classical dynamics \cite{review,Prosen}. Not much is known, however, for the temporal evolution of $F(t)$ for long times, in particular, for classically nearly integrable and mixed regular-chaotic systems.
 
An optimal platform for experiments on fidelity in low dimensional chaotic systems is provided by the quantum kicked rotor model. It is realized with cold and ultracold (Bose-Einstein condensates) atoms \cite{Raizen1999,SW2011} and with light pulses in optical fibers \cite{ulf}. The used techniques to measure fidelity range from interferometric methods in either internal atomic states \cite{oxford} or in the centre-of-mass motion of the atoms \cite{harvard} to the time reversal of the dynamics by exploiting the properties of the quantum resonant motion \cite{subfourier}.

\subsection{The quantum kicked rotor and fidelity}
\label{sec:ss3.1}

Experiments on the kicked rotor based on (ultra)cold atoms work with particles moving along a line, periodically kicked in time by an optical lattice, and possibly subject to a constant Stark or gravity field. Neglecting atom-atom interactions, the quantum dynamics is described by the following Hamiltonian in dimensionless variables (such that $\hbar=1$) \cite{fishman2002,SW2011,DGW2012}:
\begin{equation}
  \label{Hkickr}
  \hat{H}(\hat{x},\hat{p},t)\;=\;\frac{\hat{p}^2}{2}-\frac{\eta}{\tau}\; \hat{x}+ k\cos( \hat{x}) \sum_{\verb+t+\in \zZ}
\delta(t-\verb+t+\tau)\ .
\end{equation}
The kick period is $\tau$, the kick strength is $k$, and  $\verb+t+$ is a discrete time variable that counts the number of kicks.  
The additional parameter $\eta$ yields the change in momentum produced by the constant field over one kick period.
In the accelerated (by gravity) frame of reference,  the total potential is again spatially periodic, implying the conservation of quasi-momentum $\beta$. With the chosen units, $\beta$  can take on allowed values between $0$ and $1$.  
Using Bloch theory, the atom dynamics  from immediately after the $(\verb+t+-1)$-th kick to immediately after the next $\verb+t+$-th kick is then described by the unitary operator, which is derived in detail in \cite{fishman2002}:
\begin{equation}
  \hat{\cal U}_{\beta,k,\eta}(\verb+t+)\;=\;e^{-\ic k \cos(\hat{\theta})}\; e^{-\ic \tau/2(\hat{\cal N}
 + \beta +\eta\verb+t+ +\eta/2)^2}\ .
 \label{Flqt}
\end{equation}
The full evolution operator over the first $\verb+t+$ kicks is given by
\begin{equation}
  \label{U_tot}
  \hat{\cal
     U}_{\beta,k,\eta}^{\verb+t+}\;\equiv \;\hat{\cal
     U}_{\beta,k,\eta}(\verb+t-1+)\; \hat{\cal U}_{\beta,k,\eta}(\verb+t-2+)
   \dots \hat{\cal U}_{\beta,k,\eta}(\verb+1+)\; \hat{\cal U}_{\beta,k,\eta}(\verb+0+)  \,.
\end{equation}
where $\hat{\cal N}= - \ic \frac{\ud}{\ud\theta}$ is the (angular) momentum operator with periodic boundary conditions. The time-dependent  Hamiltonian that generates the quantum evolution corresponding to (\ref{U_tot}) would be
\begin{equation}
  \label{Ham1}
  \hat{\cal H}({\hat{\cal N}},\hat{\theta},\beta,t)= \frac{1}{2}
  \left(\hat{\cal N} +
      \beta +\frac{\eta}{\tau} t\right)^2+ k\cos(\hat{\theta}) \sum_{\verb+t+\in\zZ}
    \delta(t-\verb+t+\tau)\ .
\end{equation}

The temporal evolution of the fidelity can now be studied with respect to changes of one of the parameter in Eq. (\ref{Ham1}). Motivated by the experiments reported in Refs. \cite{oxford,okla1}, we consider in the following changes in the kick strength $k$ and the gravity parameter $\eta$. For a given quasimomentum $\beta$, the dynamical fidelity is defined as
\begin{equation}
  \label{fid_def1rot}
 F_\beta(k_1,k_2,\eta_1, \eta_2, \verb+t+) = \Big| \left< \hat{\cal
     U}_{\beta,k_1,\eta_1}^{\verb+t+} \psi_\beta(0) \Big|\hat{\cal
     U}_{\beta,k_2,\eta_2}^{\verb+t+} \psi_\beta(0)\right>\Big|^2 \,.
\end{equation}
In a typical experimental situations with cold atoms \cite{Ref3,qr,harvard,oxford,SW2011,okla1,okla2}, the atoms are non interacting and each of them has
in good approximation a well defined initial momentum, meaning that its quasimomentum is essentially fixed.  
This ensemble is best approximated by a statistical density operator $\hat{\rho} = \int_\beta \ud \beta \rho(\beta) \ket{\psi_\beta(0)}\bra{\psi_\beta(0)}$. Then the fidelity generalized to density operators \cite{Prosen} gives the slightly more involved formula for our system of cold atoms \cite{WB2006,DGW2012}:
\begin{equation}
  \label{fid_ens_rot}
 F(k_1,k_2,\eta_1, \eta_2,\verb+t+) =\bigg|\int_\beta \rho(\beta) \left< \hat{\cal
     U}_{\beta,k_1,\eta_1}^{\verb+t+} \psi_\beta(0) \Big|\hat{\cal
     U}_{\beta,k_2,\eta_2}^{\verb+t+} \psi_\beta(0)\right> \ud \beta \bigg|^2 \,.
\end{equation}

\subsection{Applications of fidelity to kicked cold atoms}
\label{sec:ss3.2}

\subsubsection{Quantum resonant motion and saturation of fidelity}

A series of experimental investigations has looked closely at the so-called quantum resonant motion of the quantum kicked rotor over the last fifteen years, see e.g. \cite{SW2011,qr} and references therein. This interest is mainly motivated by the type of ballistic motion with fast acceleration which can be obtained in this particular region of parameters. The quantum resonances allow one the realization of ratchet-like dynamics as well, when breaking the spatial-temporal symmetries of the kicked rotor \cite{SW2011,ratchet}. We restrict here, as most of previous works, to the main quantum resonances occurring whenever $\tau$ is not only commensurate to $2\pi$ but an integer multiple of it, i.e. $\tau=2\pi\ell$, $\ell$ integer. Then, for specially chosen values of quasi-momentum $\beta$, the energy of the rotor with this fixed $\beta$ increases quadratically as $\verb+t+\to+\infty$ \cite{sandro2003,fishman2002}.  In the presence of gravity, asymptotic quadratic growth of energy at certain values of $\beta$ is still possible close to resonant values of the kick period\cite{fishman2002}. From the theoretical point of view, these quantum resonances are convenient since we have analytical access to the wave functions \cite{sandro2003,fishman2002} and hence to analytical estimates for the development of fidelity \cite{WB2006,DGW2012}. Using pseudoclassical approximation theory \cite{sandro2003,fishman2002,AGW2009,SW2011}, this is true also for small detuning from the resonance conditions.

We review here just the result of simple analytical considerations, whose details can be found in \cite{WB2006,DGW2012}. The fidelity at a main quantum resonance is considered in two cases: first for the perturbation parameter $\Delta k=k_2-k_1$ and a fixed quasimomentum $\beta$. The fidelity is then computed from the overlap
\begin{equation}
  \left< \hat{\cal U}_{\beta,k_1,\eta}^{\verb+t+}
\psi_\beta(0) \Big|\hat{\cal U}_{\beta,k_2,\eta}^{\verb+t+} \psi_\beta(0)\right>= J_0(\Delta k
|W_{\verb+t+}|)\ ,
\label{overlap1rot}
\end{equation}
such as to yield
\begin{equation}
  \label{fid_1rot}
   F_\beta(k_1,k_2,\eta,\verb+t+)= \Big|
J_0(\Delta k |W_{\verb+t+}|)\Big|^2 \,.
\end{equation}
The argument of the zero-order Bessel function contains the following sum of phases:
\begin{equation}
 W_{\verb+t+}\equiv W_{\verb+t+}(\eta,\beta)=\sum_{r=0}^{\verb+t+-1}
e^{-\ic \pi  \ell (2\beta+1)r}e^{-\ic 2\pi \ell r \eta \verb+t++\ic\pi \ell \eta r^2
}\ .\label{Wt_1}
\end{equation}
Applying Eq. (\ref{fid_1rot}) at $\eta=0$ and resonant $\beta=\beta_{res}$ (e.g. $\beta_{res}=0$ for $\ell=2$) we have simply $W_{\verb+t+}=\verb+t+$, which leads to 
\begin{equation}
  \label{eq:fidQR}
   F_{\beta_{res}}(k_1,k_2,\eta=0,\verb+t+)= \Big|
J_0(\Delta k \verb+t+ )\Big|^2 \,.
\end{equation}
This fidelity decays asymptotically, for large times, like a power-law with exponent one \cite{WB2006}. Averaging over a full ensemble of quasimomenta in a Brillouine zone, we arrive at a fidelity which asymptotically saturates to a constant value 
\begin{equation}
	F(\Delta k, \verb+t+)	\equiv \bigg|\int_{\beta=0}^{\beta=1}\left< \hat{\cal
     U}_{\beta,k_1}^{\verb+t+} \psi_\beta(0) \Big|\hat{\cal
     U}_{\beta,k_2}^{\verb+t+} \psi_\beta(0)\right> \ud \beta \bigg|^2 
			\stackrel{t \to \infty}{\longrightarrow} \frac{1}{(2\pi)^2} \left(\int_{0}^{2\pi}
                                      J_0^2\left(\frac{\Delta k }{2\sin\alpha}\right) \ud \alpha\right)^2\,.
\end{equation}
This asymptotically exact saturation of the fidelity sets in very quickly in practice for typical experimental parameters, after $\verb+t+ > z_0/\Delta k$, where $z_0$ is the first zero of the zero order Bessel function \cite{WB2006}. An interferometric cold-atom experiment at Harvard \cite{harvard} nicely confirmed our theory showing this saturation of fidelity for various perturbations $\Delta k$ and measurements up to about 60 kicks.

In the second case, we consider the motion in a gravity field for the forward motion, whilst this field is absent in the backward motion. The kick strength does not change here. This gives for the fidelity, following a similar reasoning as done in \cite{DGW2012},
\begin{equation}\label{eq:fideta}
  F_ \beta(\eta, \verb+t+ ) = \left | \langle\psi_\beta(0)| \hat{\mathcal{U}}_{\beta, k, {\eta=0} }^{\dagger \verb+t+ } \,
 \hat{ \mathcal{U} }_{\beta, k, \eta}^{\verb+t+ } | \psi_\beta(0) \rangle \right| ^2 
= \bigg|e^{-i\phi({\beta}, {\eta}, \verb+t+)} J_0\left(\sqrt{(\verb+t+ k)^2+k^2 |W_{\verb+t+}|^2 -2 \verb+t+ k^2 \text{Re} (W_{\verb+t+}) } \right)\bigg|^2 \,.
 \end{equation}
For a single $\beta$, the additional phase $\phi({\beta}, {\eta}, t) = \ell \pi\sum_{q=0}^{t-1} \left({\beta}+q{\eta} +/ {\eta}2\right)^2$ is superfluous, yet it becomes important when considering contributions from different quasimomenta, see subsection \ref{subsecT} for this case.
			
\begin{figure}[t!]
	\begin{center}
	\includegraphics[width=0.42\linewidth]{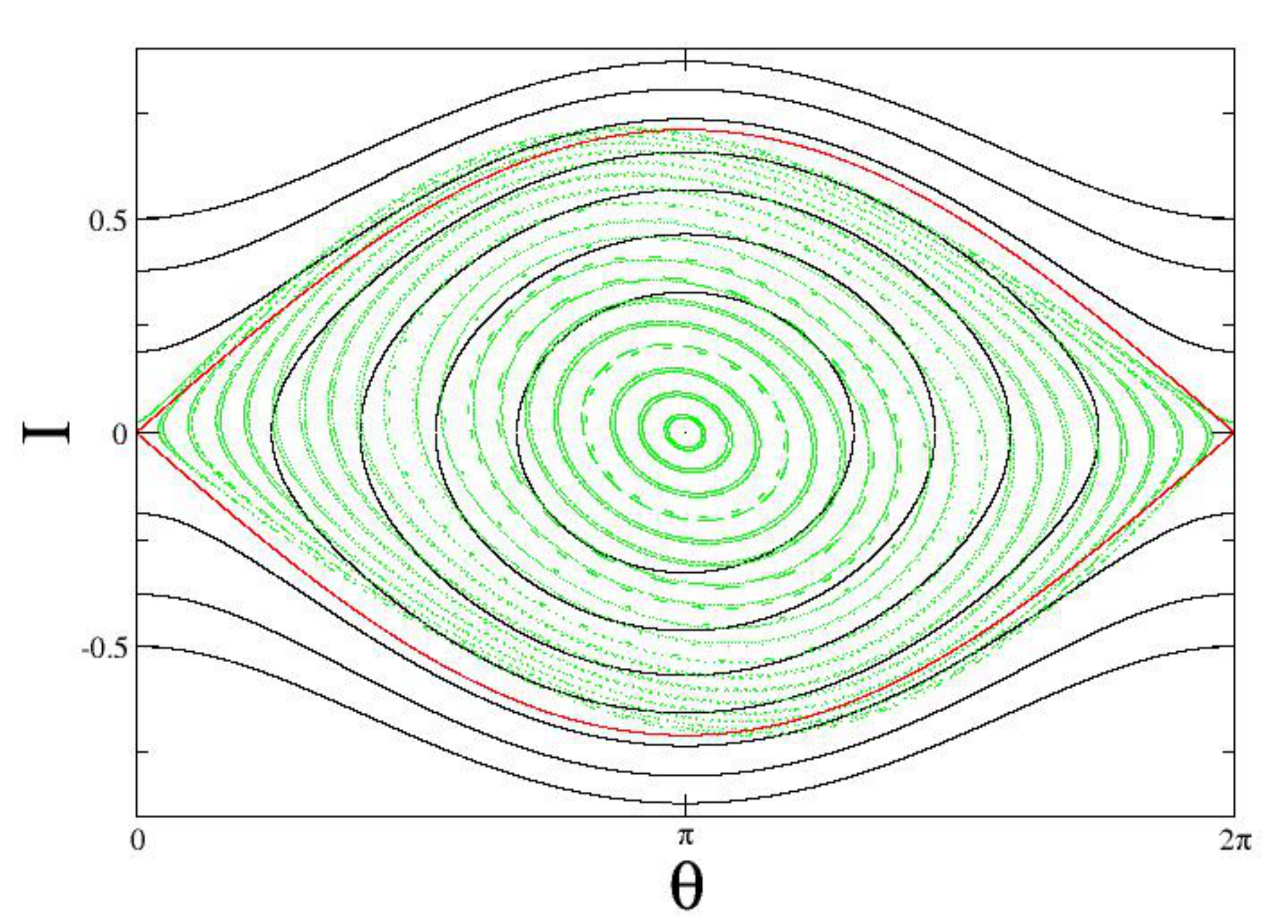}
      \includegraphics[width=0.39\linewidth]{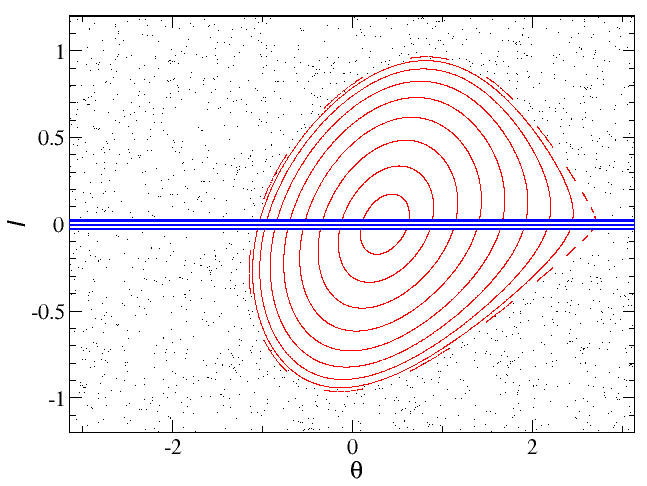}
	\caption{\label{fig:4} (left) phase space of the pseudoclassical model for $\eta=0, k|\epsilon|=0.04\pi$ (green/grey data points) as compared to its pendulum approximation \cite{wimberger2014,C1979} (solid lines, where the red line shows the separatrix of the pendulum). (right) phase space of the pseudoclassical model with gravity for the parameters $\tau=5.97,\eta=0.0257,k=1.4$. (Quasi)momentum zero corresponds to the centre of both islands on the y axis, while $\theta$ is the angle parameter of the standard map as classical limit of the kicked rotor problem. The horizon stripe in the right plot marks the typical width of an experimental quasimomentum distribution for a Bose-Einstein condensate.}
	\end{center}
\end{figure}

\subsubsection{Fidelity revivals}
\label{subsecR}

Even if we detune the kick period $\tau$ slightly from the mentioned resonance conditions, we may still be able to estimate the true quantum motion using a pseudoclassical model introduced in \cite{sandro2003,fishman2002}. In this model, the absolute value of the detuning $|\epsilon|$ plays the role of Planck's constant, and hence the theory has a semiclassical limit at exact quantum resonance.  The $\epsilon$-classical dynamics is described by the following discrete map which relates the variables $I$ and $\theta$ from immediately after the $\verb+t+$-th kick to immediately after the $(\verb+t++1)$-th one
\begin{equation}
  \label{classical}
    \begin{array}{rcl}
      I_{\verb+t++1}&=&I_{\verb+t+}+\tilde{k}\sin\theta_{\verb+t++1}+\sgn(\epsilon)\tau\eta\\
      \theta_{\verb+t++1}&=&\theta_{\verb+t+}+\sgn(\epsilon)I_{\verb+t+}\;\;\;\textrm{ mod } 2\pi
      \end{array}
      \,.
\end{equation}
Here, ${I}={J}+\sgn(\epsilon)[\pi \ell+\tau\beta+\tau\eta \verb+t+]$ and $J=|\epsilon| p$ is the rescaled momentum. Similarly, the new effective kick strength is defined by $\tilde{k}=|\epsilon| k$. Since the effective kick strength is multiplied by the small detuning in this model, the classical phase space is nearly integrable and its centre corresponds to a pendulum-like resonance island \cite{C1979,PDW2011,AGW2009,wimberger2014}, see Fig. \ref{fig:4}. In this subsection, we continue now to study the case without gravity, i.e. $\eta=0$ (see left panel of the figure). 

\begin{figure}[t!]
	\begin{center}
	\includegraphics[width=0.85\linewidth]{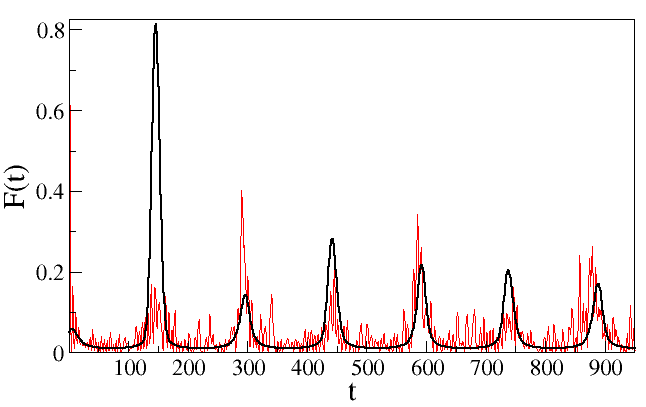}
	\caption{\label{fig:5} 
	Slightly smoothed version of the fidelity as predicted by Eq.~(\ref{eq:revival}) (thick black line) and numerical data 
	(grey/red curve) for $k_1=0.8\pi, k_2=0.6\pi$, and detuning $\epsilon=0.01$. Please note that the fidelity does start at 1 on the $y$ axis, 
	which is not visible here owing to the fast initial drop after just a few kicks. Data adapted from \cite{AGW2009}.
	}
	\end{center}
\end{figure}

Within the harmonic approximation, initial states starting at momentum close to zero will oscillate around the island, with an oscillation frequency depending on the coupling strength, $\omega_i=\sqrt{k_i |\epsilon|}$ ($i=1,2$) in this case. Hence, while the fidelity initially can decay fast at finite detuning \cite{haug2005}, it may recover if the two frequencies -- related to the two kick strength -- are commensurable. Focusing only onto resonant values of quasimomentum, the fidelity that carries over formula (\ref{eq:fidQR}) to finite detuning is
\begin{equation}
\label{eq:revival}
F_{\beta_{res}}(k_1,k_2, \verb+t+,\eta=0) \sim 
\frac{\epsilon}{2\pi} \left(| \omega_2 \cos ( \omega_1 \verb+t+ ) \sin ( \omega_2 \verb+t+)  - \omega_1 \cos ( \omega_2 \verb+t+ ) \sin ( \omega_1 \verb+t+) | \right)^{-1} \,,
\end{equation}
where the symbol $\sim$ means that this result is asymptotically true for small $\epsilon$ and large times. The predicted revivals of the fidelity are typical for integrable motion around nonlinear islands characterized by their winding numbers ($\omega_i$ measured in number of kicks) in the Poincar\'e sections. Typical results of fidelity revivals are reported in Fig. \ref{fig:5}.

\subsubsection{Dynamical tunneling}
\label{subsectionDT}

\begin{figure}[t!]
	\begin{center}
	\includegraphics[width=0.45\linewidth]{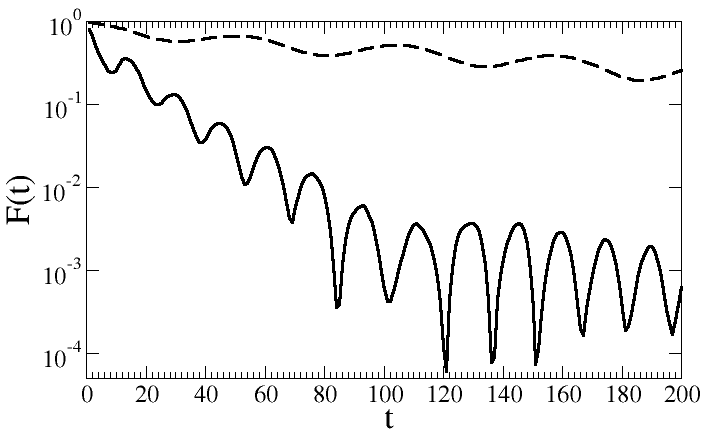}
	\includegraphics[width=0.45\linewidth]{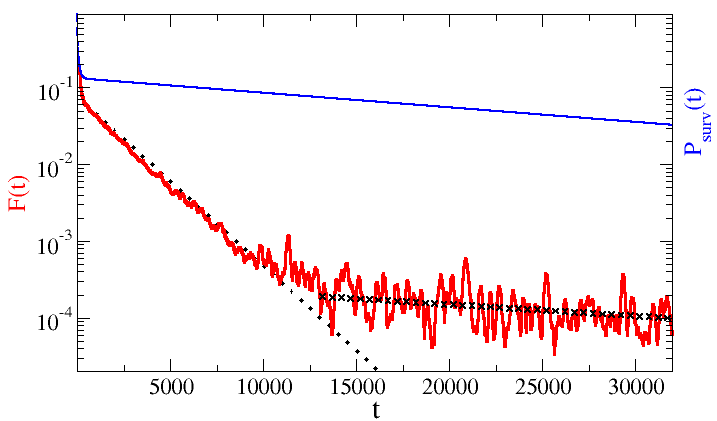}
	\caption{\label{fig:6} (left) Fidelity averaged over a broad ensemble of quasimomenta in $[0,1)$ (corresponding to a cold-atom experiment with wide
	 initial momentum distribution) for the parameters $\tau = 5.86, \eta = 0.01579\tau , k_1 = 0.8\pi, k_2 = 0.75\pi$ (dashed line) and $k_2 = 0.7\pi$ (solid line).
	(right) Comparison between the survival probability (black dots and thin blue line) and the fidelity for one fixed quasimomentum 
	(red/grey thick solid line) for $\tau=5.86$, $\epsilon=\tau-2\pi$,
      $\eta=0.01579\tau$, $\beta \approx 0.49$, $k_2=0.8\pi$, $k_1=0.7\pi$. The upper thin blue line corresponds to $k_1$ and gives the
      tunneling rates $\Gamma_1 = 5.1 \times 10^{-4}$, whereas the lower black dots corresponds to $k_2$ and gives $\Gamma_2 =
      4.4 \times 10^{-5}$. The crosses represent an exponential fit with a  rate $\Gamma = 3.5 \times 10^{-5}$ to guide the eye. Data in the right panel 
      adapted from \cite{DGW2012}.}
	\end{center}
\end{figure}

This subsection deals now with the case of a perturbation in the kick strength and one fixed value of gravity $\eta \neq 0$. All we want to say here is essentially contained in Fig. \ref{fig:6}. While initially, for short times, the fidelity suffers a fast exponential decay, this decay and the further evaluation is modulated with almost periodic oscillations. This oscillations are the better visible, the more we average over initial conditions (in this case different values of quasimomenta). The initial decay is  arising from the part of the initial wave packet that lies in the chaotic component of either of the two dynamics (defined by the two different kick strengths) in the pseudoclassical phase space, see right panel of Fig. \ref{fig:4}. While its behaviour seems to be non-uniform and strongly dependent on parameters, we can estimate the oscillation frequency by a similar argument as in the previous subsection (\ref{subsecR}). The simplest form of this estimate gives an oscillation frequency of $|\omega_1-\omega_2|+\Delta \theta$, where $\omega_i$ ($i=1,2$) were defined above and $\Delta\theta$ is the shift of the island centre in phase space induced by the change in the kick strength at finite gravity. This shift of the frequency can be a understood in the following way: for the elliptic motion in one island the trajectory is ahead, whereas for the elliptic motion in the other island the trajectory is behind, so they get closer to each other and rephase after a shorter period than in the case when both islands lie concentric. The shift, which we estimate by $\Delta\theta $, depends on the precise geometry of the two islands, and the overlapping area.

Already the data in the left panel of Fig. \ref{fig:6} shows a significant change in slope of the overall fidelity decay after some time, see the solid line at about 100 kick. We can analyze this asymptotic behaviour better focussing on an initial wave packet with just one single initial quasimomentum, see right panel of Fig. \ref{fig:6}. The fidelity is -- after the initial decay -- dominated by the parts of the wave packet which are trapped inside the islands, similar to the one shown in Fig. \ref{fig:4} (right panel). We chose our initial states in the form of Gaussian states mostly located inside one island.  The decay from the islands into the chaotic sea is a well known phenomenon known as dynamical tunneling, see in this context \cite{sheinman,okla2} and refs. therein. We will argue now that this is the main mechanism responsible of the fidelity decay at long times. The most important contribution to  fidelity  comes from the parts of each factor in the scalar product in (\ref{fid_def1rot}) that are trapped in the respective travelling island. The decay of fidelity is then in turn determined by the decay of each of these two parts,  which is determined by its respective tunneling rate
 $\Gamma_{i}$ ($i=1,2$) into the  chaotic sea. Based on this semiclassical reasoning, we came up with the following ansatz for the asymptotic decay of fidelity \cite{DGW2012}
\begin{equation}
  \label{eq:cldens}
  F(\verb+t+)\;\propto\;{\mu({\cal A}_1\backslash{\cal A}_2) e^{-\Gamma_1 \verb+t+} +
    \mu({\cal A}_2\backslash{\cal A}_1) e^{-\Gamma_2 \verb+t+} + \mu({\cal
        A}_2\cap{\cal A}_1)e^{-(\Gamma_1+\Gamma_2) \verb+t+}} \ .
\end{equation}
Here $\mu$  is the classical invariant measure of phase space sets,  ${\cal A}_i$ is the island in the phase space around the fixed point associated with $k_i$.  
In our simulations, the phase-space areas appearing in (\ref{eq:cldens})  were estimated as described  in the appendix of \cite{DGW2012}.  The quantum decay rates due to dynamical tunneling were computed numerically for each $k_i$. To do so we computed the survival probability as shown in Fig.~\ref{fig:6} by the blue dashed line. Our semiclassical formula (\ref{eq:cldens}) provides a simple yet powerful estimate for the fidelity decay since it is rather universal and applicable to a large variety of systems.

We finish this subsection by stating that first experimental measurements are on there way to detect signatures of dynamical tunneling of states prepared in quantum accelerator mode islands such as shown in the right panel of Fig. \ref{fig:4}, see ref. \cite{okla2}. The problem will remain, of course, to reach a high enough experimental stability to get access to the evolution to times longer than just a few dozens of kicks.

\subsubsection{Thermometry in ultra-cold atoms}
 \label{subsecT}

\begin{figure}[t!]
	\begin{center}
	\includegraphics[width=0.7\linewidth]{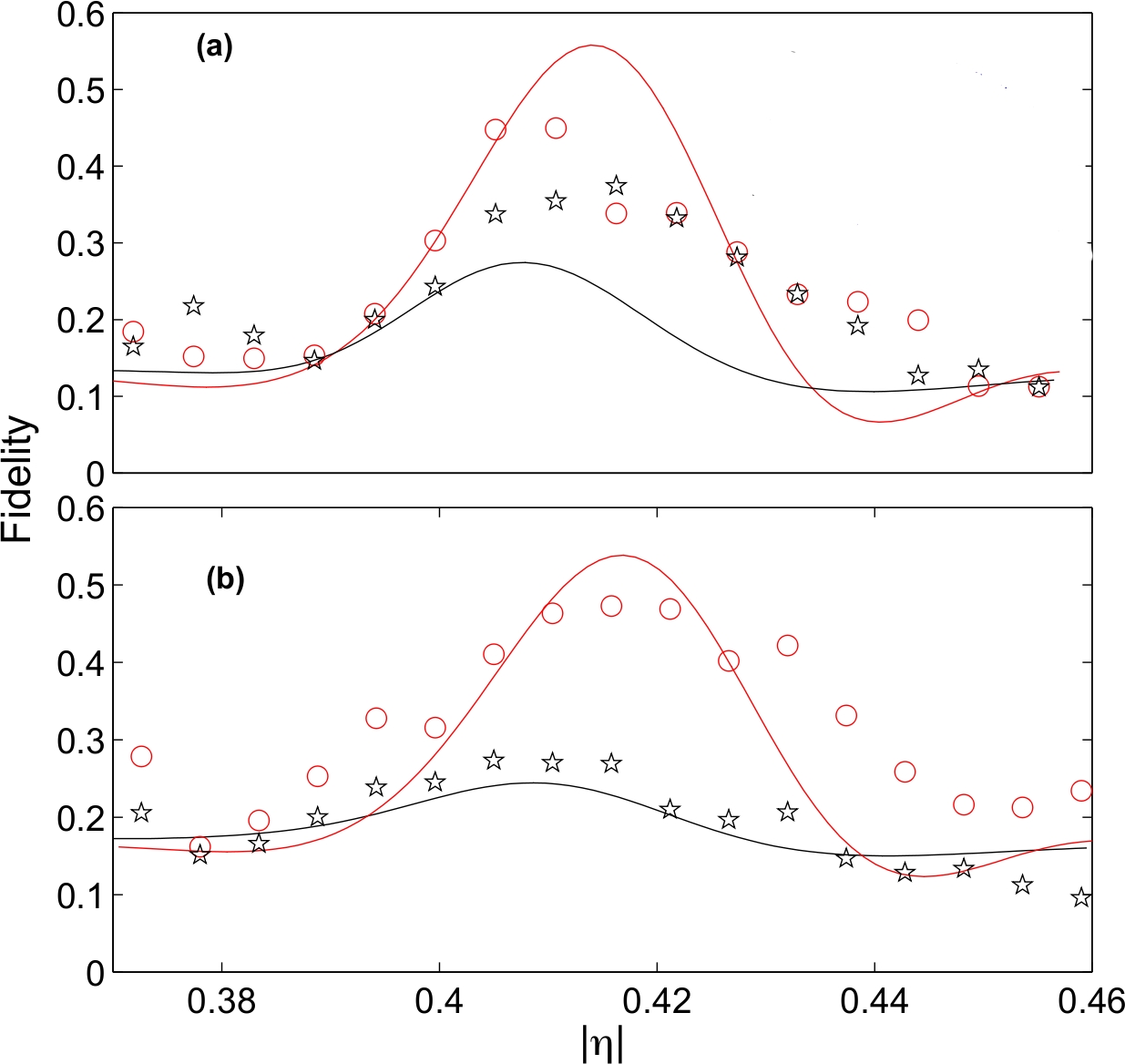}
	\caption{\label{fig:7}  Fidelities from formula (\ref{eq:fideta}) vs. $\eta$ for $\tau=4\pi, k=0.5$, taken after four kicks and averaged over a Gaussian distribution of quasimomenta with FWHM of about $\Delta \beta=0.06$ (a) and $\Delta \beta=0.07$ (b) centered around $\beta=0.5$. The solid lines are theoretical averages whilst the symbols represent experimentally measured data points. Red circles and black stars represent the fidelity with negative and positive $\eta$, respectively. We can see that the asymmetry between positive and negative gravity values, denoted $\eta_{\pm}$, increases for a broader distribution of quasimomenta allowing, in principle, the extraction of the mean kinetic energy of a cold atom ensemble from fidelity measurements. Data adapted from \cite{okla1}.}
	\end{center}
\end{figure}

\begin{figure}[t!]
	\begin{center}
	\includegraphics[width=0.35\linewidth]{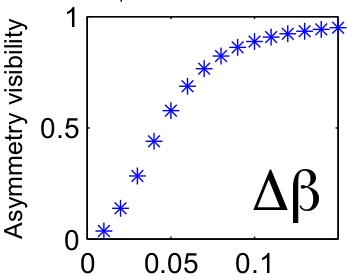}
	\caption{\label{fig:8}  Numerical simulations of an asymmetry visibility defined as $[F(\eta_-) - F(\eta_+)]/[F(\eta_-) + F(\eta_+)]$ show an almost linear scaling with the momentum width of the atomic cloud for the experimentally relevant range, see previous figure, $\Delta \beta \leq 0.08$. This linear correlation could be used to precisely define the momentum width on the $x$ axis via measurements of the asymmetry (on the $y$ axis). Figure taken from \cite{okla1}.}
	\end{center}
\end{figure}

We now turn back again at exact quantum resonance conditions, more particularly to formula (\ref{eq:fideta}) from subsection \ref{sec:ss3.2}. From the phase term $\phi({\beta}, {\eta}, t)$ in Eq.~(\ref{eq:fideta}) it can be seen that, when quasimomentum $\beta$ is non zero, then the total phase induced by different values of $\eta$ depends not only on its magnitude but also on its sign. Experiments to investigate this situation were performed at Oklahoma with a Bose-Einstein condensate of about 40 000 rubidium atoms \cite{okla1}. The atoms correspond to a distribution of quasimomenta with a FWHM of about $0.05$ in our dimensionless units. We must average Eq.~(\ref{eq:fideta}) over these initial conditions in $\beta$. The data -- guided by our theory -- clearly indicate that the asymmetric dependence of the averaged fidelity on gravity may turn out to be useful in determining externally applied accelerations and, in turn, the temperature (given by the kinetic energy) of ultracold atoms in the limit of negligible interatomic interactions. The immediate value of this experimental application is highlighted by the fact that just four kicks were used in Fig. \ref{fig:7}, making similar experiments very fast, stable and versatile. While the experimental data from Fig. \ref{fig:7} is not yet as stable as one would like for a practically useful thermometer, numerical results shown in Fig. \ref{fig:8} confirm the linear correlation between the normalized visibility of the asymmetry and the momentum width of the atomic cloud.

We conclude this section by remarking that the quantum kick rotor in the presence of gravity, as describe by the Hamiltonian (\ref{Hkickr}), is for generic values of $\eta$ not reducible to a periodically time-dependent system. Hence, we indeed need the dynamical fidelity in order to characterize the large variety of different dynamical regimes of this paradigm system of quantum chaos. A description by the static fidelity measure of section \ref{sec:2} would instead only be possible for values of $\eta$ commensurable with $\tau/|\epsilon|$, see  ref. \cite{sheinman} for details.

\section{Conclusions}
The fidelity measure has found a plethora of applications over the last sixty years, from spin and photon echos over to low-dimensional quantum chaotic systems \cite{hahn,peres, JP2001, review, Prosen}. We reviewed here two important specific fidelities, a static overlap of eigenfunctions and a dynamical overlap of generic wave functions. The former has been successfully used to detect avoided crossings in complex quantum systems, ranging from quantum phase transitions in the ground state \cite{BV2007,Ref1,Ref2} to the characterization of the chaoticity of the entire or parts of quantum spectra \cite{hiller,PLW2011,PMW2015,javier}. The much more studied dynamical fidelity has been investigated here for the paradigm system of quantum chaos, the quantum kicked rotor. We showed several applications, ranging from characterizing quasi-regular motion close to quantum resonance, over to dynamical tunneling and its indirect use as a thermometer for ultracold atomic gases. We hope that this overview will inspire further searches for interesting applications of the fidelity and generalized quantum echos, possible exploiting much more the phase dependence of the quantum mechanical overlap function for new precision experiments.

\vskip6pt

\acknowledgements{The author is grateful to Italo Guarneri and Gil Summy for stimulating our work on fidelity related to the dynamics of cold atoms. Many thanks goes to the large number of students and my former postdoc Remy Dubertrand, who all were involved in specific parts of the work reviewed here. Support by the FIL program of the Universit\`a di Parma is kindly acknowledged. The author thanks very much the organizers of the international conference {\em Echoes in Complex Systems} at the MPIPKS, in particular Arseni Goussev, for support and the kind invitation to Dresden.



\begin{thebibliography}{9}

\bibitem{hahn}
Hahn E L (1950). Spin echoes, {\it Phys. Rev.} {\bf 80} 580-594.

\bibitem{peres}
Peres A. 1984 Stability of quantum motion in chaotic and regular systems. 
\textit{Phys. Rev. A} \textbf{30} 1610-1615.

\bibitem{JP2001}
Jalabert R A, Pastawski H M. 2001. 
Environment-Independent Decoherence Rate in Classically Chaotic Systems. 
{\it Phys. Rev. Lett.}  {\bf 86} 2490.

\bibitem{review}
Jacquod P, Petitjean C. 2009. 
Decoherence, entanglement and irreversibility in quantum dynamical systems with few degrees of freedom. {\it Adv. Phys.} {\bf 58} 67-196;
Goussev A, Jalabert R A, Pastawski H M, Wisniacki D A. (2012). 
Loschmidt echo. {\it Scholarpedia} {\bf 7}(8) 11687.

\bibitem{Prosen}
Gorin T, Prosen T, Seligman T H, Znidaric M. (2006). 
Dynamics of Loschmidt echoes and fidelity decay. 
{\it Physics Reports}, {\bf 435} 33-156.

\bibitem{PLW2011}
Pl\"otz P, Lubasch M, Wimberger S. 2011. Detection of avoided crossings by fidelity.
{\it Physica A: Statistical Mechanics and its Applications} {\bf 390} 1363-1369.

\bibitem{SW2011}
Sadgrove M, Wimberger S. 2011.
A pseudo-classical method for the atom-optics kicked rotor: from theory to experiment and back.
{\it Adv. At. Mol. Opt. Phys.} {\bf 60} 315-369 (Elsevier, Amsterdam).

\bibitem{PMW2015}
Parra-Murillo C A, Madro\~nero J, Wimberger S. 2015.
Exact numerical methods for a many-body Wannier-Stark system, 
{\it Comp. Phys. Comm.} {\bf 186} 19-30.

\bibitem{javier}
Schr\"oter S, Hervieux P-A, Manfredi G, Eiglsperger J, Madro\~nero J. 2013. 
Exact treatment of planar two-electron quantum dots: Effects of anharmonicity on the complexity. 
\textit{Phys. Rev. B} {\bf 87}, 155413.

\bibitem{Ref1}
You W L, Li Y W, Gu S J 2007. Fidelity, dynamic structure factor, and susceptibility in critical phenomena. {\it Phys. Rev. E} {\bf 76}, 022101; Gu S J 2010. Fidelity approach to quantum phase transitions. {\it Int. J. Mod. Phys. B} {\bf 24} 4371

\bibitem{BV2007}
Zanardi P, Paunkovic N. 2006. Ground state overlap and quantum phase transitions. 
{\it Phys. Rev. E} {\bf 74} 031123; 
Buonsante P, Vezzani A. 2007. Ground-state fidelity and bipartite entanglement in the
Bose-Hubbard model. {\it Phys. Rev. Lett.} {\bf 98} 110601.

\bibitem{Ref2}
Campos Venuti L, Zanardi P. 2007. Quantum Critical Scaling of the Geometric Tensors. {\it Phys. Rev. Lett.} {\bf 99} 095701;
Schwandt D, Alet F, Capponi S. 2009. Quantum Monte Carlo Simulations of Fidelity at Magnetic Quantum Phase Transitions. {\it Phys. Rev. Lett.} {\bf 103} 170501; 
De Grandi C, Gritsev V, Polkovnikov A. 2010. Quench dynamics near a quantum critical point: Application to the sine-Gordon model. {\it Phys. Rev. B} {\bf 81} 224301.

\bibitem{plotz}
Pl\"otz P. 2010. Complex Dynamics of Ultracold Atoms. PhD Thesis. Heidelberg University. 
Available online at http://archiv.ub.uni-heidelberg.de/volltextserver/volltexte/2010/11123

\bibitem{JZ2005}
Jaksch D, Zoller P. 2005. The cold atom Hubbard toolbox. {\it Ann. Phys.} {\bf 315} 52.

\bibitem{haake2010}
Haake F. 2010. Quantum Signatures of Chaos, Springer-Verlag, Berlin.

\bibitem{Tom2008}
Tomadin A, Mannella R, Wimberger S. 2008. Many-body Landau-Zener tunneling in the Bose-Hubbard model. {\it Phys. Rev. A} {\bf 77} 013606.

\bibitem{Kol2004}
Kolovsky A R, Buchleitner A. 2004. Quantum chaos in the Bose-Hubbard model. {\it EPL} {\bf 68} 632

\bibitem{lubasch}
Lubasch M. 2009. Quantum Chaos and Entanglement in the Bose-Hubbard Model, 
Diploma (Master) Thesis. Heidelberg University.

\bibitem{ZK1991}
Zakrzewski J, Ku\'s M. 1991.Distributions of avoided crossings for quantum chaotic systems. 
{\it Phys. Rev. Lett.} {\bf 67} 2749.

\bibitem{wimberger2014}
Wimberger S. 2014. Nonlinear Dynamics and Quantum Chaos, Springer International Publishing, Cham.

\bibitem{hiller}
Hiller M, Kottos T, Geisel T. 2006. Complexity in parametric Bose-Hubbard Hamiltonians and structural analysis of eigenstates.
{\it Phys. Rev. A} {\bf 73} 061604(R).

\bibitem{Raizen1999}
Raizen M G. 1999. Quantum Chaos with Cold Atoms. {\it Adv. At. Mol. Opt. Phys.} {\bf 51C} 43-81 (Elsevier, Amsterdam).

\bibitem{ulf}
Peschel U. Private communication.

\bibitem{oxford} 
Schlunk S, d'Arcy M B, Gardiner S A, Cassettari D, Godun R M, Summy G S. 2003. Signatures of Quantum Stability in a Classically Chaotic System. 
{\it Phys. Rev. Lett.} \textbf{90} 054101.

\bibitem{harvard}
Wu S, Tonyushkin A, Prentiss M G. 2009. Observation of saturation of fidelity decay with an atom interferometer. {\it Phys. Rev. Lett.} {\bf 103} 034101.

\bibitem{subfourier}
McDowall P, Hilliard A, McGovern M, Gr\"unzweig T, Andersen M F. 2009. A fidelity treatment of near-resonant states in the atom-optics kicked rotor.
{\it New J. Phys.} {\bf 11} 123021;
Talukdar I, Shrestha R, Summy G S. 2010. Sub-Fourier Characteristics of a $\delta$-kicked-rotor Resonance. 
{\it Phys. Rev. Lett.} {\bf 105}, 054103; 
Ullah A, Hoogerland M D. 2011. Experimental observation of Loschmidt time reversal of a quantum chaotic system. 
{\it Phys. Rev. E} {\bf 83} 046218.

\bibitem{fishman2002}
Fishman S, Guarneri I, Rebuzzini L .2003.  A Theory for Quantum Accelerator Modes in Atom Optics.{\it J. Stat. Phys.} \textbf{110} 911.

\bibitem{DGW2012}
Dubertrand R, Guarneri I, Wimberger S. 2012. 
Fidelity for kicked atoms with gravity near a quantum resonance. {\it Phys. Rev. E} {\bf 85} 036205.

\bibitem{okla1}
Shrestha R K, Wimberger S, Ni J, Lam W K, Summy G S. 2013.
Fidelity of the quantum delta-kicked accelerator. {\it Phys. Rev. E} {\bf 87} 020902(R).

\bibitem{okla2}
Shrestha R K, Ni J, Lam W K, Summy G S, Wimberger S. 2013.
Dynamical tunneling of a Bose-Einstein condensate in periodically driven systems.
{\it Phys. Rev. E} {\bf 88} 034901.

\bibitem{qr}
d'Arcy M B, Godun R M, Oberthaler M, Cassettari D, G. S. Summy G S. 2001. 
Quantum Enhancement of Momentum Diffusion in the Delta-Kicked Rotor. {\it Phys. Rev. Lett.} {\bf 87} 074102;
Wimberger S, Sadgrove M, Parkins S, Leonhardt R. 2005.
Experimental verification of a one-parameter scaling law for the quantum and "classical" resonances of the atom-optics kicked rotor. 
{\it Phys. Rev. A} {\bf 71} 053404; Ryu C, Andersen M F, Vaziri A, d'Arcy M B, Grossman J M, Helmerson K, W. D. Phillips WD. 
2006. High-Order Quantum Resonances Observed in a Periodically Kicked Bose-Einstein Condensate. {\it Phys. Rev. Lett.} {\bf 96} 160403;
Kanem J, Maneshi S, Partlow M, Spanner M, Steinberg A. 2007. Observation of High-Order Quantum Resonances in the Kicked Rotor. 
{\it Phys. Rev. Lett.} {\bf 98} 083004.

\bibitem{Ref3}
Oberthaler M, Godun R M, d'Arcy M B, Summy  G S, Burnett K. 1999. Observation of Quantum Accelerator Modes.
{\it Phys. Rev. Lett.} {\bf 83} 4447;
Szriftgiser P, Ringot J, Deland D, Garreau G C. 2002. Observation of Sub-Fourier Resonances in a Quantum-Chaotic System. {\it Phys. Rev. Lett.}  {\bf 89} 224101.

\bibitem{WB2006}
Wimberger S, Buchleitner A. 2006. Saturation of fidelity in the atom-optics kicked rotor.
{\it Journal of Physics B: Atomic, Molecular and Optical Physics} {\bf 39} L145.

\bibitem{ratchet}
Sadgrove M, Horikoshi M, Sekimura T, Nakagawa K. 2007. Rectified Momentum Transport for a Kicked Bose-Einstein Condensate.
{\it Phys. Rev. Lett.} {\bf 99} 043002; 
Dana I, Ramareddy V, Talukdar I, Summy G S. 2008. Experimental Realization of Quantum-Resonance Ratchets at Arbitrary Quasimomenta. 
{\it Phys. Rev. Lett.} {\bf 100} 024103;
Sadgrove M, Wimberger S, Nakagawa K. 2012. Phase-selected momentum transport in ultra-cold atoms. 
{\it Eur. Phys. J. D} {\bf 66} 155; 
Shrestha R K, Ni J, Lam W K, Wimberger S, Summy G S. 2012.
Controlling the Momentum Current of an Off-resonant Ratchet. {\it Phys. Rev. A} {\bf 86} 043617;
Sadgrove M, Schell T, Nakagawa K, Wimberger S. 2013. 
Engineering quantum correlations to enhance transport in cold atoms. {\it Phys. Rev. A} {\bf 87} 013631;
Schell T, Sadgrove M, Nakagawa K, Wimberger S. 2013.
Engineering transport by concatenated maps. {\it Fluctuation and Noise Letters} {\bf 12} 1340004;
White D, Ruddell S, Hoogerland M. 2013. Experimental realization of a quantum ratchet through phase modulation.
{\it Phys. Rev. A} {\bf 88} 063603.

\bibitem{sandro2003}
Wimberger S, Guarneri I, Fishman S. 2003. Quantum resonances and decoherence for delta-kicked atoms. {\it Nonlinearity} \textbf{16} 1381.

\bibitem{C1979}
Chirikov B V. 1979. A universal instability of many-dimensional oscillator systems. {\it Phys. Rep.} {\bf 52} 263.

\bibitem{AGW2009}
Abb M, Guarneri I, Wimberger S. 2009. Pseudoclassical theory for fidelity of nearly
resonant quantum rotors. {\it Phys. Rev. E} {\bf 80} 035206.

\bibitem{PDW2011}
Probst B, Dubertrand R, Wimberger S. 2011. Fidelity of the near resonant quantum
kicked rotor. {\it Journal of Physics A} {\bf 44} 335101.

\bibitem{haug2005}
Haug F, Bienert M, Schleich W B, Seligman T H, Raizen M G. 2005. Motional stability of the quantum kicked rotor: A fidelity approach. {\it Phys. Rev. A} {\bf 71} 043803.

\bibitem{sheinman}
Sheinman M, Fishman S, Guarneri I, Rebuzzini L. 2006.  Decay of quantum accelerator modes. {\it Phys. Rev. A} \textbf{73} 052110.

\end{thebibliography}
\end{document}